\documentclass[10pt,twocolumn,twoside]{IEEEtran}
\newcommand{\comm}[1]{}
\def\xxxonly{\comm}
\def\xxxonly{ }
\def\noxxx{\comm}

\usepackage{color}

\usepackage[numbers]{natbib}\usepackage{graphicx} \usepackage{epsfig}\def\citet{\cite}

\usepackage{amssymb}
\usepackage{setspace}
\usepackage{ae} 
\usepackage{eucal} 

\newtheorem{theorem}{Theorem}
\newtheorem{lemma}{Lemma}

\newtheorem{proposition}{Proposition}

\newtheorem{definition}{Definition}

\newtheorem{example}{Example}[section]

\def\e{\varepsilon}

\def\defi{\stackrel{{\scriptscriptstyle \Delta}}{=}}

\def\a{\alpha}

\def\o{\omega}

\def\Y{{\cal Y}}

\def\w{\widehat}
\def\Ind{{\mathbb{I}}}

\def\const{{\rm const\,}}

\def\Z{{\cal Z}}

\def\H{{\cal H}}

\def\C{{\bf C}}

\def\ww{\widetilde}
\def\X{{\cal X}}

\def\oo{\bar}

\def\U{{\cal U}}

\newcommand{\be}{\begin{equation}}
\newcommand{\ee}{\end{equation}}
\newcommand{\bd}{\begin{displaymath}}
\newcommand{\ed}{\end{displaymath}}
\newcommand{\ba}{\begin{array}{ll}}
\newcommand{\ea}{\end{array}}
\newcommand{\baa}{\begin{eqnarray}}
\newcommand{\eaa}{\end{eqnarray}}
\newcommand{\baaa}{\begin{eqnarray*}}
\newcommand{\eaaa}{\end{eqnarray*}}   \font\sm=cmr10

\def\k{\kappa}

\def\ww{\tilde}

\def\H{{\cal H}}

\def\e{\varepsilon}

\def\defi{\stackrel{{\scriptscriptstyle \Delta}}{=}}

\def\a{\alpha}

\def\o{\omega}

\def\Y{{\cal Y}}

\def\w{\widehat}
\def\Ind{{\mathbb{I}}}

\def\const{{\rm const\,}}

\def\Z{{\cal Z}}
\def\ZZ{{\bf Z}}

\def\H{{\cal H}}

\def\C{{\bf C}}

\def\T{{\mathbb{T}}}
\def\TT{{\cal T}}
\def\ZZ{{\mathbb{Z}}}
\def\a{\alpha}

\def\ew{\left(e^{i\o}\right)}

\def\NN{{\scriptscriptstyle N}}
\def\BL{{\scriptscriptstyle BL}} 

\def\WW{w}
\def\wh{h}
\def\wH{H}
\def\wHH{\H}\def\EE{{\mathbb{E}}}
\date{Submitted:  April 18, 2016. Revised: September 18, 2018} 
\title{On recovering  missing  values for sequences in a pathwise setting}
\author{
Nikolai Dokuchaev }
\begin{document}
\def\break{}%
\def\brea{}
\def\breakk{}
\def\brea{\nonumber\\ }\def\breakk{\nonumber\\&&} 
\maketitle
\let\thefootnote\relax\footnote{\xxxonly{Submitted:  April 18, 2016. Revised: September 18, 2018.}
\par
The author is with  School of Electrical Engineering, Computing and Mathematical Sciences, Curtin
University,   GPO Box U1987, Perth, 6845 Western Australia\noxxx{  and also with National Research University ITMO, 197101 Russia}. }
\begin{abstract} The paper suggests a frequency criterion of  error-free recoverability of a missing value for sequences, i.e. discrete time processes,
in a pathwise setting without probabilistic assumptions. The paper establishes
 error-free recoverability for classes of square-summable sequences  with Z-transform
  vanishing   with a mild  rate at periodically located isolated points; the case of non-summable
  sequences is not excluded. The transfer functions for recovering algorithm  are presented  explicitly.
  Some robustness with respect to noise contamination is established for the suggested recovering algorithm.
\par
{\bf Key words}: data recovery, minimal Gaussian processes, pathwise criterions, frequency criterion,
discrete time, Z-transform, robustness
\end{abstract}
\section{Introduction}
A core problem of the mathematical theory of signal processing is the problem
of recovery of missing  data. For continuous data, the recoverability is associated with smoothness or analytical properties of
the processes. For
discrete time processes,  it is less obvious how to interpret analyticity; so far, these problems
were studied in a stochastic setting,
where an observed process is deemed to be  representative of an ensemble of paths with
the probability distribution that is either known or can be estimated from repeating experiments.
A classical result for stochastic stationary  Gaussian processes
with the  spectral density $\phi$ is that
a missing  single value is recoverable  with zero
error if and only if \baa
\int_{-\pi}^\pi \phi(\o)^{-1} d\o=-\infty.\label{Km} \eaa
(Kolmogorov \cite{K},
Theorem 24). Stochastic stationary Gaussian processes without this property are called {\em minimal } \cite{K}.
Criterion (\ref{Km}) was extended on stable processes \citet{Peller}  and vector Gaussian processes  \citet{Pou,Pou2}.
Clearly, (\ref{Km}) holds for all ``band-limited" processes meaning that the spectral density is vanishing on an arc of the
unit circle $\T=\{z\in\C:\ |z|=1\}$.

It is known that signals with certain restrictions on the spectrum or sparsity  feature recoverability of missing data in the pathwise setting without probabilistic assumptions.
This setting targets situations where we deal with a sole sequence that is deemed to be unique
  and such that one cannot rely on statistics  collected from observations of other similar samples.  An estimate of the missing  value
   has to be done based  on the intrinsic properties of  this sole  sequence and the observed values.
 For example,  a subsequence  sampled at sparse enough periodic points   can be removed from observations of an  oversampling sequence \cite{F95}.\index{(Ferreira (1995)).}  In the compressive sensing    setting,
 reducing of the sampling rate for finite sequences has been achieved using sparsity of signals \cite{Donoho,CJR}. The connection of
 bandlimiteness and  recoverability  from  samples was established  for
 the fractional Fourier transform
 \cite{F3}. There is also a so-called  Papoulis approach \cite{Pa} allowing to reduce the  sampling rate with additional measurements at sampling points; this approach was extended on multidimensional processes \cite{Ch1}.

There is also an approach based on the so-called Landau's phenomenon \cite{La,La2};, it was shown in  \cite{La} that there is an uniqueness  set of sampling points representing small deviations of integers for classes of functions with an arbitrarily large measure of the spectrum range.
This result was extended on functions with unbounded spectrum range and on sampling points
 allowing a convenient  explicit representation \cite{OU08}.

The paper suggests  a criterion of  error-free recoverability of sequences (discrete time processes) with spectrum degeneracy at isolated points similarly to (\ref{Km}).
  The result is based  on the approach developed for pathwise predicting  \citet{D12a,D12b,D16}, where  some predictors were derived to
  establish   error-free predictability.  In the present paper, error-free recoverability is  established for certain classes of  square-summable sequences (processes) with Z-transform
  vanishing at  a number of periodic isolated points   (Theorem \ref{ThM});  the sequences are not necessarily summable.
  The required decay rate is mild; it can be selected  as an arbitrarily low power of the distance to the nearest point of spectrum degeneracy.
   The corresponding recovering kernels are obtained and represented
  explicitly via  their transfer functions. Some robustness with respect to noise contamination is established for the suggested recovering algorithm.
A related result was obtained in \cite{D17}; however, the result obtained therein covers only the cases of sequences from $\ell_1$ with one-point spectrum degeneracy  or sequences
from $\ell_2$ that are
band-limited, i.e. with Z-transform vanishing on an  arc of the unit circle $\T$. It appears that the approach from \citet{D17} is not applicable to processes  $\ell_2\setminus \ell_1$
with spectrum degeneracy at isolated points,
since the recovery kernels used therein were non-vanishing sequences from $\ell_\infty\setminus \ell_2$.
The approach from \citet{D17} requires additional restrains on the spectrum requiring a number of derivatives for Z-transform to vanish at a point.
The approach of the present paper is quite different form the one from \citet{D17} and does not involve restrictions on the derivatives of Z-transforms; instead, it focuses on underlying processes
with spectrum vanishing at periodically located isolated points on $\T$.

 \section{Some definitions and background}
Let $\T\defi\{z\in\C:\ |z|=1\}$, and let $\ZZ$ be the set of all
integers.
\par
We denote by $\ell_r$ the set of all sequences
$x=\{x(t)\}\subset\C$, $t=0,\pm 1,\pm 2,...$, such that
$\|x\|_{\ell_r}=\left(\sum_{t=-\infty}^{\infty}|x(t)|^r\right)^{1/r}<+\infty$
for $r\in[1,\infty)$ or  $\|x\|_{\ell_\infty}=\sup_t|x(t)|<+\infty$
for $r=+\infty$.
\par
For  $x\in \ell_1$ or $x\in \ell_2$, we denote by $X=\Z x$ the
Z-transform  \baaa X(z)=\sum_{t=-\infty}^{\infty}x(t)z^{-t},\quad
z\in\C. \eaaa Respectively, the inverse $x=\Z^{-1}X$ is defined as
\baaa x(t)=\frac{1}{2\pi}\int_{-\pi}^\pi X\left(e^{i\o}\right)
e^{i\o t}d\o, \quad t=0,\pm 1,\pm 2,....\eaaa

We have that  $x\in \ell_2$ if and only if  $\|X\ew\|_{L_2(-\pi,\pi)}<+\infty$.   In addition, $\|x\|_{\ell_\infty}\le \|X\ew\|_{L_1(-\pi,\pi)}$.

We use the sign $\circ$ for convolution in $\ell_2$.

Let an integer $m\ge 0$ be given, let
$\TT_m=\{t\in\ZZ:\ |t|\le m\}$, and let $\TT_{2m}=\{t\in\ZZ:\ |t|\le 2m\}$.

Let $\wHH$ be the set  of all real sequences $\wh\in\ell_2$
such that $\wh (t)=0$ for $t\in\TT_{2m}$.

\subsection*{The setting of recovery problem}
We are interested  in the problem of recovery  values
$\{x(r)\}_{r\in\TT_m}$  from  observations $\{x(s)\}_{s:\ |s|> m}$ for $x\in\ell_2$.
More precisely,  we consider  calculation of estimates $\{\ww x(r)\}_{r\in\TT_m}$
obtained  as $\w x=h\circ x$  for some appropriate kernels $h\in\ell_2$.

This recovery operator is time invariant  and  solves  also the problem  in the time invariant setting; for any $t\in\ZZ$, the values
$\{x(t+r)\}_{r\in\TT_m}$  will be estimated from  observations $\{x(t+s)\}_{|s|>m}$
as   $\{\ww x(t+r)\}_{r\in\TT_m}$.

\par
Proposition \ref{prop2m} below shows that the recovery operators should  be based on  kernels  $h\in\wHH$.

\begin{proposition}\label{prop2m} \begin{enumerate}
\item
 If $x\in\ell_2$ and $h\in \wHH$, then, for any $r\in\TT_m$,
\baa
(h\circ x)(t+r)=\sum_{s\in\ZZ:\ |s-t|>m}h(t+r-s)x(s),
\label{hh}
\eaa
i.e. the values $\{(h\circ x)(t+r)\}_{r\in\TT_m}$ are calculated without use of $\{x(t+r)\}_{r\in\TT_m}=\{x(s)\}_{s=t+r, \ |r|\ge m}$.
\item  Let  $h\in \ell_2$ be such that
there exists $k\in\TT_{2m}$ such that $h(k)\neq 0$, i.e. $h\notin\wHH$. Then
for any   $x\in\ell_2$ there exists  $r\in\TT_m$  such that $\xi\defi r-k\in\TT_m$ and
\baaa
(h\circ x)(t+r)= h(k)x(t+\xi)\\+\sum_{s\in\ZZ:\ s\neq t+\xi}h(t+r-s)x(s).
\eaaa
\end{enumerate}
\end{proposition}
In other words,  the
values $\{(h\circ x)(t+r)\}_{r\in\TT_m}$  cannot be calculated without use of $\{x(s)\}_{s=t+r, \ |r|\le m}$
under the assumptions of  Proposition \ref{prop2m}(ii).

\begin{definition}\label{def} Let $\Y\subset \ell_2$ be a class of sequences.
\begin{itemize}
\item[(i)] We say that this class is  recoverable if
there exists a sequence $\{\wh_n(\cdot)\}_{n=1}^{+\infty}\subset
\wHH$ and \baaa \sup_{t\in\ZZ}|x(t)-\w x_n(t)|\to 0\quad
\hbox{as}\quad m\to+\infty\quad\forall x\in\Y, \eaaa where $\w x=\wh_n\circ x$.
\item[(ii)]
 We say that the class $\Y$ is  uniformly recoverable  if, for any $\e>0$, there exists $\wh(\cdot)\in \wHH$ such that \baaa \sup_{t\in\ZZ} |x(t)- \w x(t)|\le \e\quad
\forall x\in\Y, \label{predu}\eaaa where $\w x=\wh\circ x$.
\end{itemize}
\end{definition}

Since $h\circ x=\Z (H X)$, where $H=\Z h$ and $X=\Z x$, then it follows that desired recovery operators should have the following properties.
 \begin{itemize}
 \item[(a)] $h\in \H$;
 \item[(b)] $H\ew\cdot X\ew\approx X\ew$.
 \end{itemize}
We will show below that  this can be satisfied for appropriate choice of $h$ and for some wide enough classes of processes.
\section{The main results}
We will establish recoverability for sequences  with Z-transforms vanishing at some isolated  points of $\T$ with certain rate.
\subsection*{Special classes of processes}
Consider a continuous function  $W:\T\setminus\{-1\}\to (0,+\infty)$
such that $\inf_{z\in\T\setminus\{-1\}} W(z)>0$ and that, for any $\e\in(0,\pi)$, \baaa \int^\pi_{\e}W\ew d\o=+\infty.
\eaaa
\begin{example}\label{ex1}  One can select  $W\ew=(\pi-|\o|)^{-1}$, $\o\in(-\pi,\pi)$. \end{example}
\par Let $M\defi 2m+1$, i.e. $M$ is the number of elements in the set $\TT_m$.
Let
\baaa  \WW(z)\defi W(z^M),\quad z\in\T.
\label{rhok}\eaaa

Let $\X(w)$ be the class of all sequences $x\in\ell_2$ such that, for  $X=\Z x$,
\baa \int_{-\pi}^\pi \WW\ew |X\ew| d\o< +\infty.
\label{hfin}\eaa

\par
Note that  $\WW\ew \to +\infty$ as $\o\to\o_k$, where
\baaa
 \o_{k}=\frac{2\pi k-\pi}{M},\quad k\in\TT_{2m}.
\eaaa
This means that  (\ref{hfin})
 holds for ``degenerate'' processes, with $X\ew$
vanishing as $\o\to  \o_k$ with certain rate of decay  defined by $\WW$.
We call this $\WW$-degeneracy.
\par
For $\o,\w\o\in (-\pi,\pi]$, let us define a "distance" \baaa
d(\o,\w\o)\defi \min(|\o-\w\o|,|\o-(2\pi-\w\o)|).
\eaaa
Let  \baaa
D_\e\defi\left\{\o\in(-\pi,\pi]:\ \min_{k\in\TT_{2m}}d(\o,\o_k)\le \e\right\}.
\eaaa
\begin{example} Let $\X^\BL$ be the set of all $x\in\ell_2$ such that  there exists
$\e>0$ such that $X\ew|_{\o\in D_\e}=0$ (in particular, these processes are band-limited).
 Then $\X^\BL\subset\X(w)$.
\end{example}

\begin{definition}\label{defU}
 Let
 $\U(w)\subset \X(w)$   be a class of sequences. We say that this class features
 uniform $\WW$-degeneracy if, for any $x\in \U(w)$,  for $X=\Z x$,
\baaa   \int_{D_\e}\WW\ew |X\ew| d\o \to 0\quad \hbox{as}\quad \e\to 0+\brea\quad\hbox{uniformly over}\quad x\in \U(w).
 \eaaa  \end{definition}
\subsection*{Recovering kernels and recoverable sequences}
\begin{lemma}\label{lemmaH}
For $n>1$, consider  kernels constructed as $\wh_1(\cdot)=\wh_{1,n}(\cdot)=\Z^{-1}\wH_{n,1}$,  where  functions
$\wH_{1,n}$ are such that
\baaa  &\wH\ew = 1,\quad &|\o| < \pi-1/n,
\\  &\wH_{1,n}\ew = -W\ew,\quad &|\o|\in [\pi-1/n,\pi-\e_n],
\\  &\wH_{1,n}\ew = 0,\quad &|\o|\in (\pi-\e_n,\pi).
\eaaa
Here $\e_n\in (0,1/n)$ are uniquely defined such that
\baa
\int^{\pi-\e_n}_{\pi-1/n}W\ew d\o=\pi-\frac{1}{n}.
\label{en}\eaa
Then  $\wh_{1,n}\in\ell_2$ and $\wh_{1,n}(0)=0$.
\end{lemma}
\begin{example} For  $W\ew=(\pi-|\o|)^{-1}$ from Example \ref{ex1}, we have that  $\e_n=e^{1/n-\pi}/n$.
This can be calculated directly from matching
\baaa
&&-\pi+1/n=\int^{\pi-\e_n}_{\pi-1/n}W\ew d\o=\ln(\pi-\o)\bigl|_{\pi-1/n}^{\pi-\e_n}
\breakk=-\ln \e_n +\ln(1/n).
\eaaa
$\Box$
\end{example}

\begin{lemma}\label{lemmaHm}
Let $\wH_{M,n}(z)\defi \wH_{M,n}(z^M)$, $z\in\C$.
For $n>1$, consider  kernels constructed as $\wh_{M,n}(\cdot)=\Z^{-1}\wH_{M,n}$,  where  functions
$\wH_{1,n}$ is defined in Lemma \ref{lemmaH}.
Then  $\wh_{M,n}\in\ell_2$ and $\wh_{M,n}(t)=0$ if $|t|\le 2m$.
\end{lemma}

Since $\wH_{M,n}\ew= 1$ f on a large part of $(-\pi,\pi)$ for large $n\to +\infty$, 
the kernels introduced in Lemma \ref{lemmaHm}  are potential candidates for the role of
recovering kernels presented in Definition \ref{def}. The following theorem shows that these
kernels ensure required recoverability for some classes of sequences.

 \begin{theorem}\label{ThM} The following holds.
  \begin{itemize}
\item[(i)]
The class $\X(w)$ is  recoverable.
\item[(ii)] Any class $\U(w)$ featuring uniform degeneracy in the sense of Definition
\ref{defU}  is uniformly recoverable.
\item[(iii)]
The  kernels  $\wh_{M,n}$ introduced in Lemma \ref{lemmaHm}
 ensure recovering required in (i) and (ii) as $n\to +\infty$.
  For these
kernels, \baaa \sup_{t\in\ZZ}|x(t)-\w x(t)|\to 0\quad \hbox{as}\quad
n\to+\infty\quad\forall x\in\X(w). \eaaa Moreover, for any
$\e>0$, there exists $n>0$ such that \baa \sup_{t\in\ZZ}|x(t)- \w
x(t)|\le \e\quad \forall x\in\U(w). \label{pred}\eaa
Here  $\w x(t)\defi \sum_{s\in \ZZ}\wh_{M,n}(s)x(t-s).$
\end{itemize}
\end{theorem}
\subsection*{Some properties of predicting kernels}
Let us outline some essential features of kernels $h_n$.

Clearly, $h_{M,n}(t)=h_{M,n}(-t)$ for all $t\in \ZZ$.

By the choice of $H_1$, it follows that  $\sup_n\|H_{1,n}\ew\|_{L_1(-\pi,\pi)}\le 4\pi$ and $\sup_n\|h_{1,n}\|_{\ell_2}\le 2$.
Hence  $\sup_n\|h_{M,n}\|_{\ell_2}\le 2$.

Furthermore,Theorem \ref{ThM} implies  that one can decrease error by increasing $n$.
On the other hand,  we have that $\|H_{M,n}\ew\|_{L_2(-\pi,\pi)} \to +\infty$ and hence $\|h_{M,n}\|_{\ell_2}\to +\infty$ as $n\to +\infty$. This means  that the values  $|h_n(t)|$ are  decaying   as $t\to +\infty$ slower for large $n$ required for lesser recovery error. Therefore, more precise recovery would
be more impacted by data truncation and would require more observations, especially  for heavy tail inputs.

\subsection*{On the case of fast decaying inputs from $\ell_1$}
A related result based on different approach and different recovering kernels was obtained in
 in \citet{D17} for sequences $x\in\ell_1$ such that $X\ew$ is continuous in $\o=\pi$ for $X=\Z x$ together with $M-1$ derivatives in $\o$ and such that
 $X\ew$ vanishes at $\o=\pi$ together with $M-1$ derivatives. Unfortunately, the corresponding kernels $h$ obtained therein were presented by non-decaying sequences from $\ell_\infty \setminus \ell_2$.
 It seems  that the approach from \citet{D17} is not applicable to processes  $\ell_2\setminus \ell_1$. In particular, this cannot be resolved by truncation or dumping of the underlying input processes such as replacements of  $x$ by $\{x(t)r^{-|t|}\}$ with $r>1$, since the spectrum  degeneracy will not be  preserved for the amended processes.

 The approach of the present paper is quite different from the approach  \citet{D17}; in particular, it is based on different  recovering kernels $h\in \ell_2$.

\section{A numerical example}
\label{SecN}
We made some numerical experiments with recovering kernels $h_{M,n}$ defined in Lemma \ref{lemmaHm} for $m=1$, $M=3$, and $W(\o)=1/(\pi-|\o|)$.   As was mentioned above, these kernels  are real valued
even functions on $\ZZ$ such that $\| h_{M,n}\|_{\ell_2}\to +\infty$ as $n\to +\infty$.
This means that, for large $n$,  they may decay slow as $|t|\to +\infty$, which would lead to a large error
caused by inevitable data truncation.

 Figure \ref{figh}  show values of $h_{M,n}(t)$ for $m=1$,  $M=2m+1=3$ and $n=5$.

 We apply these kernel to an input  processes $x\in\H$ obtained during each Monte-Carlo simulation
 as the following.
 \begin{enumerate}
 \item A Fourier polynomial $f_1(\o)$ defined on $[-\pi,\pi]$ with $\oo N>0$ non-zero terms with independent
 random coefficients from normal distributions  was created for a given $\oo N>0$.
\item A piecewise continuous function
$f_2(\o)=\sum_{\o\in I_k}\a_k f_1(\o)$ was created for random $\a_k=\xi_k+i\zeta_k$, where $\xi_k$ and $\zeta_k$ were
selected independently from the normal distribution, and where $I_k=(-\pi+(k-1)\pi/\oo N,-\pi+k\pi/\oo N)$, $k=1,...,\oo N$.
\item We defined $X_1\ew=f_2(\o)+\overline{f_2(-\o)}$. This would ensure that  $X_1\ew=\overline{X_1\left(e^{-i\o}\right)}$, $\o\in (-\pi,\pi]$.
\item A input process $x\in\X^\BL$ was obtained using  $X_1$ as
$x=\Z^{-1}\left(\Ind_{\o\in(-\pi,\pi]\setminus D_\e}X_1\ew\right)$ for $\e\ge \e_n/3$.
\index{, where
 $D_\e=\{\o\in (-\pi,\pi]:\ \min_{k=-1,0,1}d(\o_k,\o)<\e\}$, and where $\o_{-1}=-\pi/3$,
 $\o_0=\pi$, and $\o_1=\pi/3$.}  More precisely, a finite
 set of values for input process $\{x(t)\}_{t=-N}^N$ was calculated for a given $N>0$.
\end{enumerate}
As was mentioned above, $\X^\BL\subset\X(w)$ for any choice of $W$.

It can be noted that  $n$ for the kernels $H_{M,n}$  and $\e$ for
simulated $x$  were   selected  such that $\e>1/(3n)$, i.e., $n<1/(3\e)$;  otherwise,
 the values  (\ref{hfin}) will be  too large.

We have used R software; the command {\em integrate} was used for calculation
of inverse Z-transforms for $h_{M,n}$ and $x$. \index{ It can be noted that, for the particular choice of $X_1$
used in these experiments,  $x$ could be calculated analytically for each Monte-Carlo simulation, i.e. for each   $f_1$ and $\{\a_k\}$. }   We have used $\oo N=10$.

Figure \ref{figx}  shows an example of the path for  $x$   simulated with $\e=2/9$ in two different scales.   Figure \ref{figX}  shows  the trace  $|X\ew|$ for the corresponding $X=\Z x$.

To test our algorithm for   recovery  of missing values $\{x(t)\}_{t=-1,0,1}$
from observations of   $\{x(s)\}_{s:\, |s|\ge 2}$, we calculated their estimates $\w x(t)$  using  convolution with the truncated input $x$
 \baaa
\w x(t)=\sum_{s\in\ZZ:\ |t-s|>1, |s|\le N}h_{M,n}(t-s)x(s).\eaaa
 We calculated the relative error
\baaa
E(N,n,\e)=\EE\frac{\sqrt{\frac{1}{3}\sum_{k=-1,0,1}|\w x(k)-x(k)|^2}}{\sqrt{\frac{1}{N}\sum_{t=-N}^N x(t)^2}}.
\eaaa
Here $\EE$ means average over Monte-Carlo simulations. The following
examples illustrate the impact of the choice of $N$, $n$, and $\e$, on the estimation error.
In particular, we obtained  that
 \begin{itemize}
 \item[]
$E(150,3,1/9)=0.141$, \quad
$E(150,3,2/9)=0.046$,
 \item[]
$E(300,3,1/9)=0.060$,\quad $E(300,3,2/9)=0.014$, 
  \item[]
$E(600,5,1/15)=0.039$, \quad
 $E(600,5,2/15)= 0.001$.
\end{itemize}
These examples show that, as expected, the error is decreasing as the truncation parameter  $N$ is increasing and
as the measure of the spectrum gaps $\e$  is increasing.

\section{Proofs}\label{secProof}
\par
{\em Proof of Proposition \ref{prop2m}}. For $h\in\wHH$, we have that of $r\in\TT_m$ that
\baaa
(h\circ x)(t+r)=\sum_{s\in\ZZ:\ |t+r-s|>m}h(t+r-s)x(s).
\eaaa
If $|t+r-s|>2m$ then either $t+r-s>2m$ or $t+r-s<-2m$. Since $|r|\le m$, then we have that
either $t-s>2m-r\ge m$ or $t-s<-2m-r\le -m$. In both cases, $|t-s|>m$. Hence (\ref{hh}) holds.
This completes the proof of Proposition \ref{prop2m}(i).
\par
To prove  Proposition \ref{prop2m}(ii), it suffices to observe that there exists   $r\in\TT_m$ such that $\xi=r-k\in\TT_m$.
\index{Then    $s=t+r-k$ is such that $k=t+r-s=t+xi$.}
This completes the proof of Proposition \ref{prop2m}.
$\Box$

{\it Proof of Lemma \ref{lemmaH}}. Since $\wH_{1,n}\left(e^{-i\o}\right)=
\overline{\wH_{1,n}\left(e^{i\o}\right)}$, we have that $\wh_{1,n}=\Z^{-1}\wH_{1,n}$ is real valued.
 The choice of $W$  implies  that there exists $\e_n\in (0,1/n)$ such that
(\ref{en}) holds. Further, (\ref{en}) implies that \baaa
\int_{-\pi+\e_n}^{-\pi+1/n}\wH_{1,n}\ew d\o+\int^{\pi-\e_n}_{\pi-1/n}\wH_{1,n}\ew d\o\brea =2/n-2\pi=-\int_{-\pi+1/n}^{\pi-1/n}\wH_{1,n}\ew d\o.
\eaaa
Hence $\int_{-\pi}^\pi
\wH_{1,n}\ew d\o=0$. Therefore,
\baaa \wh_{1,n}(0)=\frac{1}{2\pi}\int_{-\pi}^\pi
\wH_{1,n}\ew d\o=0.\eaaa
 This completes the proof of  Lemma \ref{lemmaH}. $\Box$
\vspace{0.5cm}

{\it Proof of Lemma \ref{lemmaHm}}. It suffices to observe that
$h_{M,n}(Mt)=h_{M,1}(t)$  and therefore $h_{M,n}(Mt)=0$ if $|t|\le M-1=2m$. $\Box$
\par
{\it Proof of Theorem \ref{ThM}}. Let $n\to +\infty$,  and let $\wH=\wH_{M,n}$  be as defined above. Let
$x\in \X(w)$, $X\defi \Z x$,  $\wh=\Z^{-1}\wH$, and
\baaa
\w x(t)\defi \sum^t_{s=-\infty}\wh (t-s)x(s). \eaaa
By
the definitions, it follows that
 $\w X\left(e^{i\o}\right)\defi \wH\left(e^{i\o}\right)X\left(e^{i\o}\right)=(\Z \w x )\ew$.
\par
\par
We have that $\|\w X\ew-X\ew\|_{L_1 (-\pi,\pi)}=I_1+I_2+I_3,$
where \baaa I_k=\int_{D_k}|\w X\ew-X\ew| d\o\brea=\int_{D_k} |(\wH\ew-1) X\ew|  d\o, \eaaa
and where $D_k=D_k(n)$, $k=1,2,3$, are defined as
\baaa &&D_1\defi (-\pi,\pi]\setminus (D_2\cup D_3),\quad\breakk
D_2\defi \cup_{k\in\TT_{2m}}\left\{\o\in(-\pi,\pi]: d(\o,\o_k)\in\left[\frac{\e_n}{m},\frac{\pi}{m n}\right]\right\},\quad\\ &&
D_3\defi \cup_{k\in\TT_{2m}}\left\{\o\in(-\pi,\pi]: d(\o,\o_k)<\frac{\e_n}{m}\right\}.  \eaaa

By the choice of $H$ and $D_1$, it follows immediately that $I_1=0$.

Let us estimate $I_2$.
 We
 have that
\baaa I_2=\int_{D_2} |(\wH\ew-1) X\ew| d\o\le \zeta_n\psi_n ,
\eaaa
where
\baaa
&&\zeta_n=\|\WW\ew X\ew
\|_{L_1(D_2)},\qquad
\breakk \psi_n=\left\|\WW\ew ^{-1}(\wH\ew-1) \right\|_{L_\infty(D_2)}.
\eaaa

We have that $\WW\ew ^{-1}\wH\ew\equiv 1$ for $\o\in D_2$. Hence $\sup_n\psi_n<+\infty$.
By the assumptions on  $X$,
\baaa
\zeta_n \to 0\quad \hbox{as}\quad n\to +\infty.
\eaaa    Hence $I_2\to 0$ as $n\to +\infty$.
\par
 Let us estimate $I_3$.
  We
 have that
\baaa &&I_3=\int_{D_3} |(\wH\ew-1) X\ew| d\o\breakk=\int_{D_3} | X\ew| d\o
\to 0\quad\hbox{as}\quad n\to+\infty.
\eaaa
 It follows that
$I_1+I_2+I_3\to 0$ for any $c>0$ and $x\in\X(w)$.  Hence $\|\w x-x\|_{\ell_\infty}\to
0$ as $n\to +\infty$ for any $x\in\X(w)$.
 This
completes the proof of statement (i).

Let us prove statement (ii). By the assumptions on $\U(w)$, it follows from the the proof above that    \baaa \|\w
X\ew-X\ew\|_{L_1(-\pi,\pi)}  = I_1+I_2+I_3\brea= I_2+I_3\to 0\quad \hbox{as}\quad n\to +\infty
\eaaa uniformly over
$x\in\U(w)$. In particular, for any $\e>0$, one can select $n$ such that
$I_2+I_3\le \e $, and this choice ensures that $\|\w x-x\|_{\ell_\infty}\le \e$.  This
completes the proof of statement (ii). It follows from the proofs above  that the
recovering kernels $\wh(\cdot)=\Z^{-1}\wH$ are such as required.
 This completes the proof of Theorem \ref{ThM}. $\Box$
 \def\NN{\eta}
 \def\nuu{\sigma}
\section{On robustness with respect to noise contamination}\label{secRob}
Let us discuss the impact of the presence
of the noise contaminating recoverable sequences.
Assume that the kernels $\wh_n$  described in Theorem \ref{ThM}  and
designed for recoverable sequences are
applied to a sequence with a noise
contamination. Let $\U(\WW)\subset \X$  be a set such as described in Definition \ref{defU}.  Let us consider an input sequence $x\in\ell_2$
such that $x=x_0+\eta$, where $x_0\in\U(\WW)$,
and where $\eta\in\ell_2\setminus \X(\WW)$ represents a
noise. Let $X=\Z x$, $X_0=\Z x_0$, and $N=\Z \eta$. We
assume that
 $\|N\ew\|_{L_1(-\pi,\pi)}=\nuu$; the parameter $\nuu\ge 0$ represents the
intensity of the noise.
\par
In the proof of Theorem \ref{ThM}, we found that, for an arbitrarily small $\e>0$, there exists $n=n(p)$ such
that \baaa
\int_{-\pi}^{\pi}|(\wH\ew-1)X_0\ew|d\o\le
2\pi\e \quad \forall x_0\in\U(w), \eaaa where $\wH=\Z \wh$. For $\w x_0=\wh\circ x_0$, this implies that
\baaa  \|\w x_0-x_0\|_{\ell_{\infty}}\le\e.\label{eps}\eaaa
\par
Let us estimate the recovery error for the case where $\nuu>0$.
For $\w x=\wh\circ x$, we have that \baaa \|\w x-x\|_{\ell_{\infty}}\le E_0+ E_{\NN},\eaaa
where\baaa &&E_0=\frac{1}{2\pi}\|(\wH\ew-1)X_0\ew|\|_{L_1(-\pi,\pi)}\le \e,\quad \breakk
E_{\NN}=\frac{1}{2\pi}\|(\wH\ew-1)N\ew|\|_{L_1(-\pi,\pi)}.
\eaaa The value $E_{\NN}$ represents the additional error caused by
the presence of unexpected high-frequency noise (when $\nuu>0$). It
follows that \baa \|\w x-x \|_{\ell_{\infty}}\le
\e+\nuu(\kappa+1),\label{yn}\eaa  where
$\kappa=\sup_{\o\in[-\pi,\pi]}|\wH\ew|$.
\par
Therefore, it can be concluded that the recovering is robust with
respect to noise contamination for any given $\e$.

It can be noted that if $\e\to 0$ then $n\to +\infty$ and $\kappa\to +\infty$. In
this case, error (\ref{yn}) is increasing for any given $\nuu>0$.
This happens when the recovering procedure is targeting too small a size of the
error for the sequences from $\X(w)$, i.e., under the assumption that
$\nuu=0$.
\par
The equations describing the dependence of $\e$ and $\k$ on $n$
could be derived similarly to estimates in \cite{D12b}, Section 6 obtained  for the predicting problem.
\section{Discussion and future development}
The  paper suggests a frequency  criterion of  error-free recoverability of a finite set of missing values
in pathwise deterministic setting in the spirit of the Kolmogorov's criterion of minimality
for stochastic Gaussian
stationary processes \citet{K}.  The paper  suggests a robust recovering algorithm for classes of these sequences with periodic spectrum gaps.

Theorem \ref{ThM} gives a criterion of recoverability that, for a special case where $M=1$, reminds  the classical Kolmogorov's  criterion (\ref{Km}) of minimal recoverability  formulated for the spectral densities \citet{K}.
However, the degree of similarity is quite limited. For instance, if a stationary Gaussian process has the
spectral density $\phi(\o)\le \const\cdot (\pi-|\o|)$, $\o\in (-\pi,\pi]$, then, according to criterion (\ref{Km}), this process is  minimal \citet{K},
i.e. a single missing value of this process is non-recoverable. On the other hand, Theorem \ref{ThM} applied with $M=1$
 and $W(\o)=\WW\ew =(\pi-|\o|)^{-1}$  implies that   the class $\X(w)$ is recoverable. In particular,  this class includes all sequences $x\in\ell_2$ such that $|X\ew|\le \const \cdot (\pi-|\o|)^{\a}$, where $X=\Z x$ and $\o\in (-\pi,\pi]$, and where $\a>0$ can be arbitrarily small.  A similar example is given in \citet{D17} for the case of summable processes from $\ell_1$.
We leave this analysis for the future research.

There are other open questions. In particular, it is unclear if it is possible to  obtain pathwise necessary  conditions of recoverability in this pathwise setting based on Z-transform.

 \def\sm{}
\begin{figure}[ht]
\centerline{\epsfig{figure=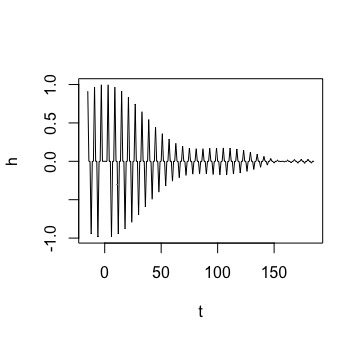,width=9cm,height=6.0cm}}
\vspace{-0.5cm}
\centerline{\epsfig{figure=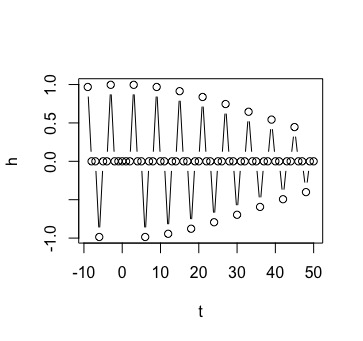,width=9cm,height=6.0cm}}
\caption[]{\sm The kernel $h_{M,n}(t)$ for $m=1$, $M=3$, and $n=5$, for $t=0,1,...,200$ and
$t=0,1,...,50$ (below). }
\label{figh}
\end{figure}
\begin{figure}[ht]
 \centerline{\epsfig{figure=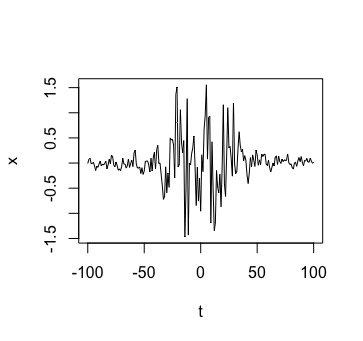,width=9cm,height=6.0cm}}
 \vspace{-1cm}
\centerline{\epsfig{figure=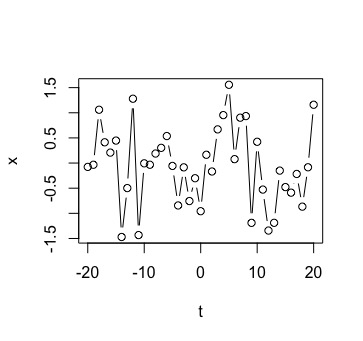,width=9cm,height=6.0cm}}
\caption[]{\sm The path  $x$ for  $t=-300,...,300$ and for   $t=-30,...,30$ (below) simulated with $\e=2/15$. }
\label{figx}
\end{figure}
\begin{figure}[ht]
 \centerline{\epsfig{figure=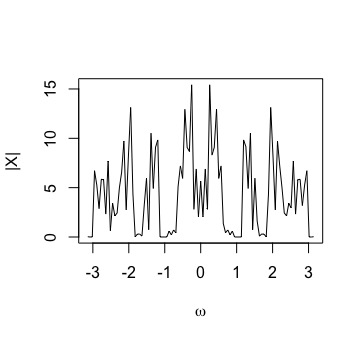,width=9cm,height=6.0cm}}
\caption[]{\sm The trace of $|X\ew|$ for the simulated path $x$ shown in Figure \ref{figx}.
It  can be seen that the spectrum vanishes at the neighbourhood of points  $\pi$ and $\pm \pi/3$.}
\label{figX}
\end{figure}
  \end{document}